%% file: arxiv.tex
\documentclass{article}
\usepackage{ijcai23}

\usepackage{times}
\usepackage{soul}
\usepackage{url}
\usepackage[hidelinks]{hyperref}
\usepackage[utf8]{inputenc}
\usepackage[small]{caption}
\usepackage{graphicx}
\usepackage{amsmath}
\usepackage{amsthm}
\usepackage{booktabs}
\usepackage[switch]{lineno}
\usepackage{array}
\usepackage{flushend}
\usepackage{subcaption}

\usepackage{algorithm}
\usepackage[noend]{algpseudocode}
\algrenewcommand\algorithmiccomment[1]{\hfill\textcolor{blue}{\commentsymbol{} #1}}

\usepackage{subfiles}
\usepackage{amssymb}
\usepackage{tikz,pgffor}
\usepackage{ifthen}
\usetikzlibrary{arrows,backgrounds,calc}
\usetikzlibrary{automata,positioning,arrows,shapes,math,arrows.meta,decorations.pathmorphing}
\tikzstyle{min}=[thick,circle,draw,minimum size=1.4em,inner sep=0em,text centered]
\tikzstyle{dec}=[circle,draw,fill,minimum size=.8ex,inner sep=0em]
\usepackage{scalerel}

\input{macros}

\definecolor{ourviolet}{RGB}{72, 35, 117}
\definecolor{ourgreen}{RGB}{33, 145, 141}
\definecolor{ourblue}{RGB}{53, 95, 141}
\hypersetup{
	final,
	colorlinks,
	breaklinks=true,
	linkcolor=ourblue,
	citecolor=ourblue,
	urlcolor=ourblue,
}

\title{Synthesizing Resilient Strategies\\ for Infinite-Horizon Objectives in Multi-Agent Systems}

\author{
David Kla\v{s}ka\and
Anton\'{\i}n Ku\v{c}era \and 
Martin Kure\v{c}ka \and
V\'{\i}t Musil \and  
Petr Novotn\'{y} \and 
Vojt\v{e}ch \v{R}eh\'{a}k
\affiliations
Masaryk University, Brno, Czech Republic
\emails
tony@fi.muni.cz
}

\begin{document}

\maketitle

\begin{abstract}
We consider the problem of synthesizing resilient and stochastically stable strategies for systems of cooperating agents striving to minimize the expected time between consecutive visits to selected locations in a known environment.
A strategy profile is \emph{resilient} if it retains its functionality even if some of the agents fail, and \emph{stochastically stable} if the visiting time variance is small.
We design a novel specification language for objectives involving resilience and stochastic stability, and we show how to efficiently compute strategy profiles (for both autonomous and coordinated agents) optimizing these objectives.
Our experiments show that our strategy synthesis algorithm can construct highly non-trivial and efficient strategy profiles for environments with general topology.
\end{abstract}

\import{sections}{intro}
\import{sections}{model}

\import{sections}{algorithms}
\import{sections}{experiments}

\import{sections}{conclusion}

\appendix
\import{sections}{graphs}

\section*{Acknowledgments}

Research was sponsored by the Army Research Office and was accomplished under
Grant Number W911NF-21-1-0189. 

\noindent
\textit{Disclaimer.}\quad The views and conclusions contained in this document are those of the authors
and should not be interpreted as representing the official policies, either
expressed or implied, of the Army Research Office or the U.S.\ Government. The
U.S.\ Government is authorized to reproduce and distribute reprints for
Government purposes notwithstanding any copyright notation herein.

Martin Kure\v{c}ka received funding from the European Union’s Horizon Europe program under the Grant Agreement No.\ 101087529. Petr Novotný is supported by the Czech Science Foundation grant GA23-06963S.

\import{sections}{appendix}

\import{sections}{references}
\end{document}

%% file: macros.tex
\newcommand{\Exp}{\mathbb{E}}
\newcommand{\calA}{\mathcal{A}}

\newcommand{\calC}{\mathcal{C}}
\newcommand{\calE}{\mathcal{E}}
\newcommand{\calF}{\mathcal{F}}

\newcommand{\calH}{\mathcal{H}}

\newcommand{\calG}{\mathcal{G}}

\newcommand{\Time}{\textit{Time}}
\newcommand{\Reach}{\mathit{Reach}}

\newcommand{\Prob}{\operatorname{Prob}}

\newcommand{\Dist}{\operatorname{Dist}}

\newcommand{\Var}{\textit{Var}}

\newcommand{\Succ}{\mathit{Succ}}

\newcommand{\A}{\mathcal{A}}

\newcommand{\MT}{\textit{ET}}

\newcommand{\MTV}{\textit{VT}}
\newcommand{\Ag}{\textit{Ag}}
\newcommand{\pp}[3]{$({\color{red}#1},{\color{blue}#2}#3)$}

\usepackage{xspace}
\makeatletter
\DeclareRobustCommand\onedot{\futurelet\@let@token\@onedot}
\def\@onedot{\ifx\@let@token.\else.\null\fi\xspace}

\makeatother

%% file: sections/intro.tex
\section{Introduction}
\label{sec-intro}

In multi-agent path planning, the terrain is modeled as a directed graph where the nodes correspond to possible agents' positions and the edges represent admissible moves. The moving plan (strategy) can be either \emph{coordinated} or \emph{autonomous} for each agent. 

A classical problem of cooperative multi-agent path planning is minimizing the time lag between consecutive visits to certain locations. Variants of this unbounded horizon planning problem are studied in connection with persistent data gathering, remote software protection, periodic maintenance (where the service nodes are distributed in space), or surveillance/patrolling problems where mobile agents strive to detect possible intrusions at protected locations. The existing approaches to strategy synthesis 
can be classified into three main types: (A) splitting the nodes into disjoint subsets, assigning agents to these subsets, and computing a special strategy for each agent/subset; (B) assigning the same strategy to all agents with different initial positions; (C) specific techniques applicable to restricted topologies, such as open/closed perimeter.

The first approach is sensitive to agent failures, because each node is visited only by one agent. If such an agent is not willing or able to report the attack (in which case we call the agent \emph{faulty}), the node covered by this agent becomes susceptible to an attack. (To account for the worst case we assume that such a faulty behaviour cannot be detected.) The second approach results in strategies that are more resilient but generally less efficient. The third approach produces good results, but only for selected topologies. Finally, most of the existing algorithms 
compute only \emph{deterministic} strategies, even in scenarios where \emph{randomized} strategies achieve better performance (see the example below).  

\paragraph{Our Contribution} We design a class of objective functions sufficiently rich to express preferences on the \emph{maximal time} needed for visiting a certain subset of location from every reachable configuration and the level of \emph{resilience} with respect to agent failures. Since we allow for \emph{randomized} solutions (strategy profiles), the objective functions can also specify the required \emph{stochastic stability} of the constructed solution to prevent large deviations from its expected performance. Furthermore, we design \emph{efficient strategy synthesis algorithms},
and we show that these algorithms can automatically discover sophisticated and well-performing solutions even for general instances with irregular topology. In some cases, the discovered solutions \emph{outperform} the best existing results. The algorithm also rediscovers sophisticated solutions for special topologies that were designed \emph{manually}.

More concretely, we introduce a class of \emph{fault-tolerant recurrent visit (FTRV)} objectives built upon atoms of the form $\MT(v,f)$ and $\MTV(v,f)$, where $v$ is a \emph{target node} and $f$ is the number of \emph{faulty} agents. Here,
\begin{itemize}
\item $\MT(v,f)$ is the maximal \emph{expected time} for visiting target $v$ by a non-faulty agent;
\item $\MTV(v,f)$ is the maximal \emph{variance} of the time for visiting target $v$ by a non-faulty agent.
\end{itemize}
In both cases, the maximum is considered over all reachable configurations and all possible selections of $f$ faulty agents (we refer to Section~\ref{sec-model} for precise semantics).

A \emph{FTRV objective function} is a function of the form 
\begin{equation}
    \alpha_1 \cdot \max \calE_1 + \cdots + 
                         \alpha_m \cdot \max \calE_m \,,
\end{equation}
where every $\alpha_i$ is a positive \emph{weight}, and every $\calE_i$ is a finite set of terms built over numerical constants and atoms of the form $\MT(v,f)$ and $\MTV(v,f)$ using differentiable functions.

Hence, a FTRV objective function is a weighted sum of requirements referring to the maximal \emph{expected time} for visiting a target node by a non-faulty agent and the corresponding variance. The goal is to \emph{minimize} this function by constructing strategies ``implementing'' all of these requirements simultaneously.

Our strategy-synthesis algorithm computes randomized finite-memory strategies for a given number of agents. The memory states represent some information about the sequence of previously visited nodes. In the autonomous case, each agent has its own memory, and makes decisions independently of the other agents. In the coordinated case, all agents share the same memory and make their decisions ``collectively''. In both cases, the strategies are constructed from randomly chosen strategies by gradient descent, and the algorithm improves all strategies \emph{simultaneously}. This ensures that the agents tend to cooperate even in the autonomous case. 

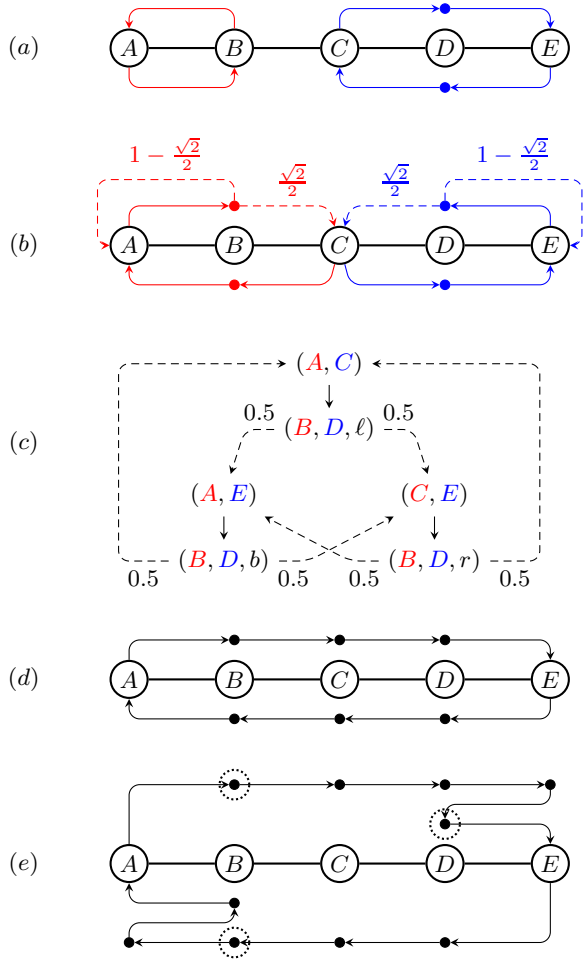
\begin{figure}[t]\centering
\begin{tikzpicture}[x=2cm, y=2.5cm, scale=0.7,font=\small]
\foreach \X/\l in {0/A,1/B,2/C,3/D,4/E}{%
   \foreach \Y/\L/\LL in {0/a/a, -1.5/b/b,-4.8/c/d, -6.2/d/e}{%
      \node [min] (\L\X) at (\X,\Y) {$\l$};
      \ifthenelse{\NOT \X=0}%
          {\draw[thick] (\L\X) -- (\L\the\numexpr\X-1\relax) ;}%
          {\node at ($(\L\X) +(-1,0)$) {$(\LL)$};}
   }
} 
\node[dec,color=blue] (bl) at ($(a3) +(0,-.3)$) {};
\node[dec,color=blue] (bu) at ($(a3) +(0,.3)$) {};
\draw[-stealth,rounded corners,color=red] (a0) -- ($(a0) +(0,-.3)$) -- ($(a1) + (0,-.3)$) -- (a1); 
\draw[-stealth,rounded corners,color=red] (a1) -- ($(a1) +(0,.3)$) -- ($(a0) + (0,.3)$) -- (a0); 
\draw[stealth-,rounded corners,color=blue] (a2) -- ($(a2) +(0,-.3)$) -- (bl);
\draw[stealth-,rounded corners,color=blue] (bl) -- ($(a4) + (0,-.3)$) -- (a4); 
\draw[stealth-,rounded corners,color=blue] (a4) -- ($(a4) +(0,.3)$) -- (bu);
\draw[stealth-,rounded corners,color=blue] (bu) -- ($(a2) + (0,.3)$) -- (a2); 
\node[dec,color=red]  (rl) at ($(b1) +(0,-.3)$) {};
\node[dec,color=red]  (ru) at ($(b1) +(0,.3)$) {};
\node[dec,color=blue] (bl) at ($(b3) +(0,-.3)$) {};
\node[dec,color=blue] (bu) at ($(b3) +(0,.3)$) {};
\draw[stealth-,rounded corners,color=red] (b0) -- ($(b0) +(0,-.3)$) -- (rl); 
\draw[-stealth,rounded corners,color=red] (b0) -- ($(b0) +(0,.3)$) -- (ru); 
\draw[stealth-,rounded corners,color=red] (rl) -- ($(b2) +(-.1,-.3)$) -- (b2); 
\draw[stealth-,rounded corners,color=blue] (b4) -- ($(b4) +(0,-.3)$) -- (bl); 
\draw[stealth-,rounded corners,color=blue] (bl) -- ($(b2) +(.1,-.3)$) -- (b2); 
\draw[-stealth,rounded corners,color=red,densely dashed] (ru) -- ($(ru) +(0,.2)$) -- node[above]{$1-\frac{\sqrt{2}}{2}$} ($(b0) + (-.3,.5)$) |-  (b0);
\draw[-stealth,rounded corners,color=red,densely dashed] (ru) -- ($(b1) +(.2,.3)$) -- node[above]{$\frac{\sqrt{2}}{2}$} ($(b2) + (-.1,.3)$) -- (b2);
\draw[-stealth,rounded corners,color=blue,densely dashed] (bu) -- ($(bu) +(0,.2)$) -- node[above]{$1-\frac{\sqrt{2}}{2}$} ($(b4) + (.3,.5)$) |-  (b4);
\draw[-stealth,rounded corners,color=blue,densely dashed] (bu)-- node[above]{$\frac{\sqrt{2}}{2}$} ($(b2) + (.1,.3)$) -- (b2);
\draw[-stealth,rounded corners,color=blue] (b4) -- ($(b4) +(0,.3)$) -- (bu);
\node[dec]  (u1) at ($(c1) +(0,.3)$) {};
\node[dec]  (u2) at ($(c2) +(0,.3)$) {};
\node[dec]  (u3) at ($(c3) +(0,.3)$) {};
\node[dec]  (l1) at ($(c1) +(0,-.3)$) {};
\node[dec]  (l2) at ($(c2) +(0,-.3)$) {};
\node[dec]  (l3) at ($(c3) +(0,-.3)$) {};
\draw[-stealth,rounded corners] (c0) |- (u1);
\draw[-stealth,rounded corners] (u1) -- (u2);
\draw[-stealth,rounded corners] (u2) -- (u3);
\draw[-stealth,rounded corners] (u3) -| (c4);
\draw[-stealth,rounded corners] (c4) |- (l3);
\draw[-stealth,rounded corners] (l3) -- (l2);
\draw[-stealth,rounded corners] (l2) -- (l1);
\draw[-stealth,rounded corners] (l1) -| (c0);
\node[dec]  (uu1)  at ($(d1) +(0,.6)$) {};
\node[dec]  (uu2)  at ($(d2) +(0,.6)$) {};
\node[dec]  (uu3)  at ($(d3) +(0,.6)$) {};
\node[dec]  (u3)   at ($(d3) +(0,.3)$) {};
\node[dec]  (uu4)  at ($(d4) +(0,.6)$) {};
\node[dec]  (ll0)  at ($(d0) +(0,-.6)$) {};
\node[dec]  (ll1)  at ($(d1) +(0,-.6)$) {};
\node[dec]  (l1)  at ($(d1)  +(0,-.3)$) {};
\node[dec]  (ll2)  at ($(d2) +(0,-.6)$) {};
\node[dec]  (ll3)  at ($(d3) +(0,-.6)$) {};
\draw[-stealth,rounded corners] (d0) |- (uu1);
\draw[-stealth,rounded corners] (uu1) -- (uu2);
\draw[-stealth,rounded corners] (uu2) -- (uu3);
\draw[-stealth,rounded corners] (uu3) -- (uu4);
\draw[-stealth,rounded corners] (uu4) -- ($(uu4) +(0,-.15)$)  -| (u3);
\draw[-stealth,rounded corners] (u3) -|  (d4);
\draw[-stealth,rounded corners] (d4) |- (ll3);
\draw[-stealth,rounded corners] (ll3) -- (ll2);
\draw[-stealth,rounded corners] (ll2) -- (ll1);
\draw[-stealth,rounded corners] (ll1) -- (ll0);
\draw[-stealth,rounded corners] (ll0) -- ($(ll0) +(0,.15)$)  -| (l1);
\draw[-stealth,rounded corners] (l1) -| (d0);

%
%

\coordinate (a) at (1.9,-2.4);
\node at ($(b0) +(-1,-1.5)$) {$(c)$};
\node (AC)  at (a) {\pp{A}{C}{}};
\node (BDl) at ($(a) +(0,-.5)$)  {\pp{B}{D}{,\ell}};
\node (AE)  at ($(a) +(-1,-1)$) {\pp{A}{E}{}};
\node (CE)  at ($(a) +(1,-1)$)  {\pp{C}{E}{}};
\node (BDs) at ($(a) +(-1,-1.5)$) {\pp{B}{D}{,b}};
\node (BDr) at ($(a) +(1,-1.5)$)  {\pp{B}{D}{,r}};
\draw[-stealth,rounded corners] (AC) -- (BDl);
\draw[-stealth,rounded corners] (AE) -- (BDs);
\draw[-stealth,rounded corners] (CE) -- (BDr);
\draw[-stealth,rounded corners,densely dashed] (BDl) -- node[above]{$0.5$} ($(BDl) +(-.8,0)$)  -- (AE);
\draw[-stealth,rounded corners,densely dashed] (BDl) -- node[above]{$0.5$} ($(BDl) +(.8,0)$)  -- (CE);
\draw[-stealth,rounded corners,densely dashed] (BDs) -- node[below]{$0.5$} ($(BDs) +(.8,0)$)  -- (CE); 
\draw[-stealth,rounded corners,densely dashed] (BDs) -- node[below]{$0.5$} ($(BDs) +(-1,0)$)  |- (AC); 
\draw[-stealth,rounded corners,densely dashed] (BDr) -- node[below]{$0.5$} ($(BDr) +(-.8,0)$)  -- (AE); 
\draw[-stealth,rounded corners,densely dashed] (BDr) -- node[below]{$0.5$} ($(BDr) +(1,0)$)  |- (AC); 
\draw[densely dotted, thick] (uu1) circle (8pt);
\draw[densely dotted, thick] (u3) circle (8pt);
\draw[densely dotted, thick] (ll1) circle (8pt);
\end{tikzpicture}
\caption{The deterministic solution of (a) for two agents is outperformed by the autonomous and coordinated randomized solutions of (b) and (c). Resilient solutions for two and three agents are shown in (d) and (e).}
\label{fig-strat}
\end{figure}

\paragraph{Example} We illustrate FTRV objectives and the functionality of our strategy synthesis algorithm on the graph of Fig.~\ref{fig-strat}(a) with five nodes $V = \{A,B,C,D,E\}$ arranged into a line where traversing each edge takes $1$ time unit (this models an open perimeter with five locations at regular intervals). Even for this simple instance, our algorithm constructs solutions \emph{outperforming} the best known strategies, and also rediscovers some results presented in previous works. Since these observations are important, we explain them in greater detail.   

Let us first consider the problem of constructing strategies for two reliable agents (red and blue) such that the maximal expected time for visiting each node is as small as possible and both strategies are stochastically stable to a chosen degree. 
This is expressed by a FTRV objective\footnote{The objective aims at minimizing the maximal value of the sum $\MT(v,0) + \kappa \cdot \sqrt{\MTV(v,0)}$ over all reachable configurations, see Section~\ref{sec-model}.} 
\begin{equation}
  \textbf{minimize}\
	\max\big\{ \MT(v,0) + \kappa \cdot \sqrt{\MTV(v,0)} \mid v \in V\big\}.
\label{obj-time}
\end{equation}
Here, $\kappa \geq 0$ is a constant  ``punishing'' the standard deviation
$\sqrt{\MTV(v,0)}$ (a smaller deviation is enforced by a larger~$\kappa$). We start with the case when $\kappa {=} 0$, i.e., we optimize just the maximal expected time for visiting a node.

One trivial solution is to follow the aforementioned approach~(A), split the nodes among the agents, and construct two trivial ``cycling'' strategies of Fig.~\ref{fig-strat}(a). The maximal $\MT(v,0)$ is then equal to $3$, regardless of the initial agents' positions (the blue agent needs $3$ time units to visit $C$ when it is in $D$ and moves to $E$ in the next step). Actually, this is the \emph{best} outcome achievable by any \emph{deterministic} solution\footnote{This can be proven as follows. For the sake of contradiction, assume there is a determininistic solution s.t.{} the maximal $\MT(v,0)$ is~$2$. Then $C$ must be visited by some agent after $k \geq 1$ time units. Hence, this agent visits nodes $X$,$C$,$Y$ after $k{-}1$,$k$,$k{+}1$ time units, where $X,Y \in \{B,D\}$. This means that the other agent must visit \emph{both} $A$ and $E$ in the time interval $k{-}1$,$k$,$k{+}1$, which is impossible.}
 (the existing algorithms construct only deterministic strategy profiles). However, our algorithm discovers \emph{better} solutions, both in the autonomous and the coordinated case. 

In the autonomous case, our algorithm computes (a rational approximation of) the solution of Fig.~\ref{fig-strat}(b). Both agents use two memory states, because the decision taken in $B$ (or~$D$) depends on whether the red (or blue) agent came from the node on the left or right. The initial configuration is $(\textcolor{red}{A},\textcolor{blue}{D_\ell})$, i.e., the red agent is in $A$, and the blue agent is in~$D$, behaving as if it came from~$C$. The maximal $\MT(v,0)$ is~$1{+}\sqrt{2} \approx 2.41$, attained for, e.g., $\MT(C,0)$ in the configuration $(\textcolor{red}{A},\textcolor{blue}{D_\ell})$. Hence, this solution outperforms any deterministic solution in the \emph{expected} performance. However, one can still argue that the 
probability of visiting $C$ from $(\textcolor{red}{A},\textcolor{blue}{D_\ell})$ in $4$ or more time units is positive, while the deterministic solution of Fig.~\ref{fig-strat}(a) does not suffer from this deficiency.


In the coordinated case, our algorithm discovers the coordinated strategy of Fig.~\ref{fig-strat}(c). The initial configuration is $(\textcolor{red}{A},\textcolor{blue}{C})$, and then the agents collectively move to successor configurations in the indicated way. The three shared memory states $\ell,r,b$, indicate whether the left/right/both agent(s) choose the next move randomly in $(\textcolor{red}{B},\textcolor{blue}{D})$. Note that the agents' decisions in $(\textcolor{red}{B},\textcolor{blue}{D},b)$ are \emph{not} independent. The maximal $\MT(v,0)$ is equal to~$2$, which is \emph{optimal}\footnote{To see this, realize that when $C$ is visited by some agent, then the other agent must be located in a node of either $\{A,B,C\}$ or $\{C,D,E\}$, and hence at least two time units are needed to visit $E$ or $A$, respectively.}, i.e., there is no solution such that the maximal $\MT(v,0)$ is less than~$2$. Furthermore, every node is visited in at most $3$ time units with probability $1$ from every reachable configuration. Hence, this solution outperforms any deterministic solution in \emph{expected} performance, achieving the same \emph{worst-case} performance.

For $\kappa > 0$, the constructed solution in the autonomous case ``trades'' performance for stability, i.e., the maximal $\MT(v,0)$ increases for increasing $\kappa$. For a sufficiently large $\kappa$, we obtain a deterministic strategy where the maximal $\MT(v,0)$ equals~$3$ and the maximal $\MTV(v,0)$ is~$0$. In the coordinated case, we obtain the same coordinated strategy of Fig.~\ref{fig-strat}(c) even for $\kappa > 0$, because this strategy is actually rather stable (the maximal $\MTV(v,0)$ is equal to $1$).

All of the above solutions suffer from \emph{low resilience}. If one agent fails, some nodes will not be covered at all, i.e., the maximal $\MT(v,1)$ is $\infty$. To obtain a resilient solution, we use a FTRV objective
\begin{equation}
\label{obj-time-stable}
\begin{split}
  \textbf{minimize} \quad 
		& \max\big\{ \MT(v,0) \mid v \in V\big\}
			\\
  + & \alpha \cdot \max\big\{ \MT(v,1) \mid v \in V\big\}.
 \end{split}
\end{equation}
Here, we wish to minimize \emph{both} the maximal $\MT(v,0)$ and the maximal $\MT(v,1)$. For a sufficiently large $\alpha$, our algorithm produces the solution of Fig.~\ref{fig-strat}(d). Both agents execute the \emph{same} deterministic cycle through all nodes of length $8$, and the initial configuration is $(\textcolor{red}{A},\textcolor{blue}{E})$. Hence, the maximal $\MT(v,0)$ is $3$, and the maximal $\MT(v,1)$ is $7$. Note that a sufficiently large $\alpha$ naturally leads to avoiding randomization, because the best strategy for \emph{one} agent is deterministic. In this case, we obtain the same solution as existing algorithms following the aforementioned approach~(B). 

For \emph{three} agents and objective~\eqref{obj-time-stable}, our algorithm discovers the solution of Fig.~\ref{fig-strat}(e). All agents execute the same deterministic cycle of length $12$. The initial configuration is indicated by the three dotted circles (two agents are initially in $B$, but for different memory states representing different visits to $B$ along the cycle). The maximal $\MT(v,0)$ is $1$, and the maximal $\MT(v,1)$ is $5$. This solution closely resembles the solution for \emph{continuous time} multi-agent patrolling the open perimeter designed \emph{manually} in \cite{KS:Multi-patrol-strategies-TCS}. Hence, our algorithm rediscovers the design pattern of \cite{KS:Multi-patrol-strategies-TCS}  in the discrete time setting, but for somewhat different reason. The solution of \cite{KS:Multi-patrol-strategies-TCS} is constructed to achieve the best coverage of nodes by a deterministic solution. We aim at solution optimizing the coverage both for $0$ and $1$ failing agent, which for suitable $\alpha$ leads to a deterministic strategy prioritizing the maximal $\MT(v,0)$ over the maximal $\MT(v,1)$. 

Let us note that for certain values of $\alpha$, our algorithm produces \emph{randomized} solutions. For example, when $\alpha$ is suitably small, the solution is similar to the one obtained for the FTRV objective~\eqref{obj-time} where $\kappa = 0$, and slightly ``adjusted'' so that all agents visit all nodes repeatedly. 



\subsection{Related Work}
\label{sec:related-work}

Strategy synthesis for multi-agent systems is a rich research area that has been deeply studied for decades \cite{Shoham:book,Wooldridge:book,Dixon:book}. The \emph{finite-horizon} path planning problems for (multi-)agent systems are among the most researched subjects in mobile robotics (see, e.g., \cite{Choset:book,LaValle:book}). Recent technological advances motivate the study of \emph{infinite-horizon} path planning problems where the agents (robots, humans, software processes) perform an uninterrupted task such as persistent data-gathering \cite{STBR:temporal-planning-IJRS}, remote software protection \cite{BLM:Remote-SW-Protection,CT:Codebender-IEEESoftware,CMMN:distributed-tamper-detection}, or patrolling \cite{HZHH:multi-robot-patrol-survey,ARSTMCC:multi-patrolling-survey,PR:multi-patrolling-survey}. Some of the classical finite-horizon planning problems, such as the vehicle routing problem or the generalized traveling salesman problem \cite{TV:book}, are solved by constructing collections of deterministic cycles that can be followed arbitrarily long. Hence, they can be seen as solutions to the corresponding infinite-horizon planing problems in situations where the underlying environment does not change. 

The existing strategy synthesis techniques are mainly based on analyzing the structural properties of the underlying graph (such as splitting the graph into smaller units assigned to individual agents that are subsequently solved by special methods), constructing a uniform strategy followed by all agents, or special techniques applicable to restricted topologies such as lines or circles. Since the terrain is known, agents are cooperative, and the planning horizon is infinite, the techniques for multi-agent strategy learning (see, e.g., \cite{GD:multiagent-learning-survey}) have not been found  particularly advantageous in this context. Our algorithm is based on differentiable programming and gradient descent, inspired by the approach used in \cite{KKMR:Regstar-UAI} for single-agent adversarial patrolling. 

Randomized strategies have been used mainly in adversarial patrolling based on Stackelberg equilibria \cite{SFAKT:Stackelberg-Security-Games,YKKCT:Stackelberg-Nash-security} to reduce agents' predictability. Otherwise, randomization has been mostly avoided in infinite-horizon path planning, apparently for several reasons: randomized strategies are not apt for human agents (drivers, police squads, etc.), they are harder to construct, and their capability for delivering better performance (as demonstrated in the above example) is not immediately apparent. 

Resilience to agent failures has so far been studied for (non-adversarial) surveillance and deterministic strategies. In \cite{HK:robustness-multirobot-ICRA}, a robust solution is obtained by constructing a cycle in the underlying graph followed by all agents with shifted initial positions. It is observed that longer cycles visiting some nodes repeatedly may improve performance (the strategy of Fig.~\ref{fig-strat}(e) constructed by our algorithm has the same property). In \cite{CGKKKT:corrupted-patrolmen-Algorithmica}, the strategy synthesis for $n$ agents out of which precisely $f$ are faulty is studied, again for deterministic strategies. This corresponds to the FTRV objective 
\begin{equation}
 \textbf{minimize}\quad \max\{\MT(v,f) \mid v \in V\}.
\end{equation}
However, considering deterministic strategies is insufficient for achieving optimal results (even if $f=0$), as demonstrated in Fig.~\ref{fig-strat}(b)(c). In particular, deterministic strategies cannot use the ``entagled'' randomized choice performed by the coordinated strategy of Fig.~\ref{fig-strat}(c), which is crucial for decreasing the maximal $\MT(v,0)$ to~$2$.

General specification languages for infinite-horizon objectives in multi-agent systems are mostly based 
on temporal logics (see, e.g., \cite{HW:logics-multiagent-survey} for an overview. A formula of such a logic specifies desirable properties of trajectories, and the constructed strategies are deterministic. The idea of trading performance for stochastic stability had been studied for systems with one agent, where the underlying objectives are specified as mean payoff functions \cite{BCFK:performance-stability-JCSS} or recurrent reachability criteria \cite{KKMR:RecReach-IJCAI}.      

To the best of our knowledge, the results of this paper are the first attempt to solve the optimization problem for complex objective functions ``balancing'' the requirements on the expected time for visiting configurations, resilience, and stochastic stability of the constructed solutions for $n$~agents. Contrasting to previous works on multi-agent strategy synthesis for infinite-horizon objectives, our algorithm computes \emph{randomized} solutions (autonomous or coordinated), achieving strictly better performance than deterministic solutions produced by previous works.  

%
%
%

%% file: sections/model.tex
\section{Mathematical Model}
\label{sec-model}
We assume familiarity with basic notions of probability theory (expected value, variance, etc.) and Markov chain theory. A finite \emph{Markov chain} is represented as a pair $M=(S,\Prob)$ where $S$ is a finite set of states and $\Prob: S\times S \to [0,1]$ is a \emph{stochastic matrix} such that $\sum_{s' \in S}\Prob(s,s') = 1$ for every $s \in S$. For a given state $s$ and a subset $F \subseteq S$, we use  $\Exp[\Time(s{\to}F)]$ to denote the expected length of a trajectory from $s$ to a state of $F$, and $\Var[\Time(s{\to}F)]$ to denote the corresponding variance. A state of $M$ is \emph{reachable} if it is visited from a given \emph{initial} state with positive probability. For a finite set $A$, we use $\Dist(A)$ to denote the set of all probability distributions over~$A$.

\subsection{Environment}
\label{sec-environment}
The environment is modeled as a directed graph $\calG = (V,E)$ where the vertices $V$ correspond to locations visited by the agents and the edges $E \subseteq V \times V$ model admissible agents' moves between the locations. For simplicity, we assume that traversing each edge takes one time unit (general traversal times can be modeled by inserting auxiliary vertices). 
%
%
For the rest of this section, we fix an environment $\calG = (V,E)$.
 
%

\subsection{Strategies for Autonomous Agents}
In the autonomous case, every agent $A_i$ uses its private finite set $M_i$ of \emph{memory states} to store some information about the sequence of previously visited vertices. The next move of $A_i$ is selected randomly according to the currently visited vertex and the current memory element. 

More precisely,  
a \emph{moving strategy} for $A_i$ is a function \mbox{$\sigma_i : V {\times} M_i \to \Dist(V {\times} M_i)$}. We require that whenever $\sigma_i(v,m)$ selects $(v',m')$ with positive probability, then $(v,v') \in E$. 

A \emph{strategy profile} for agents $A_1,\ldots,A_n$ is a tuple $\sigma = (\sigma_1,\ldots,\sigma_n)$, where every $\sigma_i$ is a moving strategy for $A_i$. 
A~\emph{configuration} is a tuple 
$[(v_1,m_1),\ldots,(v_n,m_n)]$ describing the current vertex and the current memory state of every agent. From a given \emph{initial} configuration, the agents start to execute their moving strategies simultaneously and independently. Thus, the agents  proceed from one configuration to another. The next configuration is reached in one time unit, and it is selected randomly in the expected way. 

More precisely, we define a Markov chain $M_\sigma$ where the states are the configurations, and for all configurations $c = [(v_1,m_1),\ldots,(v_n,m_n)]$ and $c' = [(v'_1,m'_1),\ldots,(v'_n,m'_n)]$, we put $\Prob(c,c') = \prod_{i=1}^n \sigma_i(v_i,m_i)(v'_i,m'_i)$. 


\subsection{Strategies for Fully Coordinated Agents}

In the fully coordinated case, the information about the histories of agents' moves is stored in a ``global'' memory $M$ consisting of finitely many states, and a \emph{coordinated strategy} for $n$ agents is a function $\pi : V^n {\times} M \to \Dist(V^n {\times} M)$. Hence, the next move of every $A_i$ depends on the current positions of all agents and the current state of the global memory. Here, a \emph{configuration} becomes a tuple of the form $(v_1,\ldots,v_n,m)$, and the Markov chain $M_\pi$ over the configurations is defined in the straightforward way.

\subsection{FTRV Objectives}
\label{sec-objectives}

The class of \emph{fault-tolerant recurrent visit (FTRV)} objectives is built upon \emph{atoms} of the form $\MT(v,f)$ and $\MTV(v,f)$, where $v \in V$ is a \emph{target node} and $f \geq 0$ is the number of \emph{faulty agents}. We start by explaining the semantics of these two atoms.

Let $\mu$ be a strategy profile or a coordinated strategy for $n$~agents where $n > f$ (i.e., at least one agent is not faulty). For all $v \in V$ and $\calA \subseteq \{A_1,\ldots,A_n\}$, let $\calC(v,\calA)$ be the set of all configurations of $M_\mu$ where at least one agent of $\calA$ is located in $v$. Furthermore, let $\Reach$ be the set of all reachable configurations of $M_\mu$, and let $\Ag[f]$ be the set of all $\calA \subseteq \{A_1,\ldots,A_n\}$ such that $\calA$ contains precisely $n{-}f$ agents. The \emph{$(\mu,c,\A)$-value of $\MT(v,f)$}, where $c \in \Reach$ and $\calA \in \Ag[f]$,
is defined as follows:
\begin{equation}
  \MT^{\mu,c,\A}(v,f) ~=~  \Exp[\Time(c{\to}\calC[v,\calA])]
\end{equation}
Hence, $\MT^{\mu,c,\A}(v,f)$ is the expected time for visiting target $v$ from the configuration $c$ by an agent of $\A$.

Similarly, the \emph{$(\mu,c,\A)$-value of $\MTV(v,f)$}, denoted by $\MTV^{\mu,c,\A}(v,f)$, is defined as the variance of the time for visiting target $v$ from $c$ by an agent of $\A$, i.e.,
\begin{equation}
  \MTV^{\mu,c,\A}(v,f) ~=~  \Var[\Time(c{\to}\calC[v,\calA])].
\end{equation}

A \emph{term} is an expressions $t$ built over numerical constants and atoms using differentiable functions such as multiplication or addition (each atom in $t$ may use different $v$ and~$f$). The \emph{$(\mu,c,\A)$-value of $t$}, denoted by $t^{\mu,c,\A}$, is obtained by substituting each atom in $t$ with its $(\mu,c,\A)$-value and evaluating the resulting expression. Furthermore, we define $t^{\mu} = \max_{c \in \Reach}\max_{\calA \in \Ag[f]} t^{\mu,c,\A}$.

A \emph{FTRV objective function} is an expression $U$ of the form
\begin{equation}
	U ~\equiv~ \alpha_1 \cdot \max \calE_1 + \cdots + 
                         \alpha_m \cdot \max \calE_m
\end{equation}
where every $\alpha_i$ is a positive \emph{weight}, and every $\calE_i$ is a finite set of terms.  The \emph{$\mu$-value of $U$}, denoted by $U^\mu$, is defined as 
$\sum_{i=1}^m \alpha_i\cdot \max \{t^\mu \mid t \in \calE_i\}$.

A \emph{FTRV objective} is an objective of the form $\textbf{minimize}\ U$, where $U$ is a FTRV objective function.

\subsubsection{Examples}

Simple examples of FTRV objectives are given in Section~\ref{sec-intro}. Here we show how to express some of the unbounded-horizon path planning objectives studied for multi-agent systems in previous works. 

A widely accepted effectiveness measure for deterministic strategy profiles in robotics is \emph{idleness}, i.e., the maximum time between successive visits of each node.  Let $\sigma$ be a deterministic strategy profile achieving a finite idleness $\mathcal{I}$. Then, every node of $V$ is visited infinitely often by some agent, and the longest time elapsed between successive visits to a given $v \in V$ is equal to $\MT(v,0)+1$.
Hence, the problem of minimizing idleness is expressible as the FTRV objective
\begin{equation}
 \begin{split}
 \textbf{minimize} \quad \max
 		& \{\MT(v,0) \mid v \in V\}
			\\
 	+ & \alpha \cdot \max \{\MTV(v,0) \mid v \in V\}.
 \end{split}
\end{equation}
The first summand is the idleness (the ``$+1$'' can be safely removed from all terms, because the resulting objective is equivalent), and the second summand  ``enforces'' determinism with a suitable weight~$\alpha$. Since robotic agents can easily implement randomized strategies, more efficient solutions can be obtained by using objective~\eqref{obj-time}, as demonstrated in Section~\ref{sec-intro}. 
 
In adversarial patrolling, a malicious \emph{attacker} observes the agents and aims to initiate an attack (e.g., set a fire at a chosen node) maximizing the damage proportional to the time of discovering the attack and the vulnerability of the node. The goal is to minimize the damage caused by an \emph{optimal} attack. Since the attack may be initiated right after all agents start moving to the next configuration, the worst expected time for discovering an attack at $v$ is $\MT(v,0)+1$ and not just $\MT(v,0)$. Hence, the patrolling objective can be expressed as 
\begin{equation}
 \textbf{minimize} \quad  \max \{w_v \cdot(\MT(v,0){+}1) \mid v \in V\},
\label{obj-patrol}
\end{equation}
where $w_v$ is a constant representing the vulnerability (importance) of $v$.
Objective~\eqref{obj-patrol} can be further refined by adding requirements on stochastic stability and/or resilience. Such refinements have not been studied in previous works.  


Finally, let us note that the strategy profiles and coordinated strategies constructed by our synthesis algorithms are \emph{ergodic}, i.e., for every $\mu$ there exists a unique \emph{limit frequency} of visits to every reachable configuration~$c$, denoted by $\mathbb{F}^\mu(c)$. If we additionally fix a probability distribution $\calF$ on $\Ag[f]$ such that $\calF(\A)$ is the probability that the agents of $\A$ are correct under the condition that precisely $f$ agents are faulty, we can rigorously define the \emph{long-run average} \mbox{$\mu$-value} of every term $t$  by 
\begin{equation}
    \textit{Avg}^\mu(t) ~=\! \sum_{c \in \Reach} \mathbb{F}^\mu(c) \cdot\! \sum_{\A \in \Ag[f]} \calF(\A) \cdot t^{\mu,c,\A}
\end{equation}
and enrich our language of FTRV objectives with the \textit{Avg} operator. For the sake of simplicity, we keep our current setting.

%% file: sections/algorithms.tex
\section{Strategy Synthesis Algorithm}
\label{sec-algorithms}
 
In this section, we describe our strategy synthesis algorithm for autonomous strategy profiles and coordinated strategies. In principle, these are two different algorithms, but their functionality is similar and it possible to describe both of them at once. The main difference is the sets of parameters representing a strategy profile $\sigma$ and a coordinated strategy~$\pi$. The Markov chains $M_\sigma$ and $M_\pi$ are constructed differently (see Section~\ref{sec-model}), but our algorithm processes them in the same way. 

For the rest of this section, we fix a graph $\calG = (V,E)$. We collectively refer to strategy profiles and coordinated strategies as \emph{solutions}.

Our algorithm is based on differentiable programming and gradient descent, and it performs the standard optimization loop shown in Algorithm~\ref{alg:optim}. We start by identifying a set of real-valued parameters representing a solution. 
\begin{itemize}
 \item For an autonomous strategy profile, for every agent $A_i$ and every $(v,m) \in V \times M_i$, we need $|\Succ(v) \times M_i|$ parameters to represent the distribution $\sigma_i(v,m)$, where $\Succ(v)$ is the set of immediate successors of~$v$. The size of each $M_i$ is a hyper-parameter of our algorithm. 
 \item For a coordinated $\pi$, we need $|M| \cdot \prod_{i=1}^n|\Succ(v_i)|$ parameters
 to represent the distribution  $\pi(v_1,\ldots,v_n,m)$.
\end{itemize}
These parameters are initialized to random values sampled from \textit{LogUniform} distribution so that we impose no prior knowledge about the solution. Then, these values are transformed into probability distributions using the standard \textit{Softmax} function, obtaining the corresponding solution $\mu$.

The crucial ingredient of Algorithm~\ref{alg:optim} is a procedure for \emph{evaluating} a given FTRV objective function $U$ for $\mu$ (see Section~\ref{subsec-algo-evaluation}). This procedure allows to compute $U^\mu$, and also the gradient of $U$ at the point corresponding to $\mu$ by automatic differentiation. After that, we update the point representing the current $\mu$ in the direction of the steepest descent. The intermediate solutions and the corresponding values of $U$ are stored, and the best solution found within $\textit{Step}$ optimization steps is returned. Our implementation uses
\textsc{PyTorch} framework~\cite{PyTorch} and its automatic differentiation with \textsc{Adam} optimizer~\cite{Adam}).

\begin{algorithm}[t]
\small
\caption{Solution synthesis}
\label{alg:optim}
\begin{algorithmic}[t]
\State ${\rm SolutionParameters}\gets {\it RandomInit}(V, E)$
\For{$i\in \{1,\ldots,{\rm Steps}\}$} 
        \State ${\mu} \gets {\it Softmax}({\rm SolutionParameters})$
        \State ${U^\mu} \gets {\it Evaluate}({\mu})$
        \State ${\nabla U(\mu)} \gets {\it Gradient}({\mu})$
        \State ${\rm SolutionParameters~~} {+}{=} {\it ~~Step}({\nabla U(\mu)})$
        \State \textbf{Save} ${U^\mu}, {\mu}$\vspace{0.5ex}
\EndFor
\Return ${\mu}$ with the least $U^\mu$
\end{algorithmic}
\end{algorithm}

\subsection{Evaluating Solutions}
\label{subsec-algo-evaluation}

Let us fix a FTRV objective function $U$ and a solution $\mu$. Recall the definition of the Markov chain $M_\mu$ presented in Section~\ref{sec-model}. For all $c,d\in M_\mu$, we use $\mu(c,d)$ to denote the probability of the transition from $c$ to $d$ (i.e., the value $\Prob(c,d)$, where $\Prob$ is the stochastic matrix of $M_\mu$). 

Let $\calH$ be the underlying directed graph of $M_\mu$, where the vertices are configurations, and $(c,d)$ is an edge of  $\calH$ iff $\mu(c,d)>0$.  First, we apply the Tarjan's algorithm \cite{Tarjan:SCC-decomp-SICOMP} to find all bottom strongly connected components (BSCCs) of $\calH$. Note that for each BSCC $B$ of $\calH$, the value of $U$ is the same for all initial configurations $c\in B$. Moreover, the value of $U$ for an initial configuration $c$ not belonging to any BSCC cannot be lower than the value of $U$ obtained for an initial configuration belonging to a BSCC reachable from~$c$. Hence, it suffices to compute the value of $U$ for each BSCC~$B$ separately, and choose the initial configuration so that it belongs to the best BSCC (observe that this BSCC can be seen as an ergodic Markov chain; this explains the remarks at the end of Section~\ref{sec-objectives}). 

So, let us fix a BSCC $B$, a target $v \in V$, a number of faulty agents $f < n$, and $\calA \subseteq \{A_1,\ldots,A_n\}$ such that $|\calA|=n-f$.
For the sake of brevity, let $T_c$ stand for $\Time(c{\to}\calC[v,\calA])$. We show how 
to compute $\Exp[T_c]$ and $\Var[T_c]$ for all $c\in B$.

If $B\cap \calC[v,\calA]=\emptyset$, then 
$\Exp[T_c]=\infty$ for all $c\in B$,
and the BSCC $B$ is disregarded, because the agents are unable to cover $v$. Otherwise,
we create a system of linear equations over variables $(X_c)_{c\in B}$. For each $c\in B$,
we have the equation
\begin{equation*} \label{E:expectation}
X_{c}  = 
\begin{cases}
0 & \text{if $c \in \calC[v,\calA]$,}\\
1+\sum_{d \in B} \mu(c,d) \cdot X_{d}
& \text{otherwise.}
\end{cases}
\end{equation*}
Since $B$ is a BSCC, it follows from standard results of Markov chain theory (see, e.g.~\cite{Norris:book})
that this system has a unique solution, equal to
$(\Exp[T_c])_{c\in B}$.

The computation of $\Var[T_c]$ is similar. We create a system of linear equations over variables $(X_c)_{c\in B}$. For every \mbox{$c\in B$},
we have the equation
\begin{equation*} \label{E:variance}
X_{c}  = 
\begin{cases}
0 & \text{if $c \in \calC[v,\calA]$,}\\
1+\sum_{d \in B} \mu(c,d) \cdot (2\Exp[T_d]+X_{d})
& \text{otherwise.}
\end{cases}
\end{equation*}
Again, this system has a unique solution, equal to
$(\Exp[{T_c}^2])_{c\in B}$.
Finally, we obtain that $\Var[T_c]=\Exp[{T_c}^2]-(\Exp[T_c])^2$
for each $c\in B$.

Having evaluated all atoms, the value of $U$ in $B$ is computed in the straightforward way. Since the terms may contain only differentiable functions, we can still use automatic differentiation to compute the gradient of~$U$.


%% file: sections/experiments.tex
\section{Experimental Results}
\label{sec-experiments}

\newcommand{\pholder}{\textcolor{red}{XXX}}
\newcommand{\tocheck}[1]{\textcolor{red}{#1}}
\newcommand{\kpath}[1]{$P_{#1}$}
\newcommand{\gridn}[1]{\((4\times4)\)-\textit{grid}\(#1\)}
\newcommand{\trigr}{$\Delta$}
\newcommand{\stdev}{\sqrt{VT}^{\ \max}}
\newcommand{\obj}{E}
\newcommand{\meanE}{\MT^{\ \max}}
\newcommand{\meanER}{\MT_R^{\ \max}}

We focused on the following questions pertaining to our algorithm: (A) whether it is able to find or approximate optimal solutions in (smaller) instances where optimality can be easily verified manually; (B) how it scales w.r.t. increasing size of the input; (C) how does the incorporation of resilience and stochastic stability into the objectives and of memory into the agent's state affect its performance and behaviour; and (D) how does it perform on graphs with various topologies.

\subsection{Benchmarks}
We experimented with several benchmarks. The first is the \textit{open perimeter benchmark} with 5 nodes discussed in Section~\ref{sec-intro}, denoted by \kpath{5}. We also consider generalizations to \kpath{k} (open perimeters of length~$k$) for increasing odd values of~$k$ to address question~(B) above. The next benchmark we consider is a \( 4\times 4\) grid in which several edges were removed in a way preserving connectedness. This creates an irregular topology through which we address question~(D). We consider several graphs of this form.  Finally, we consider the ``triangle'' \trigr{} benchmark. This can be seen as a closed perimeter (a circle) of length 6 with a ``shortcut'' in the center. The exact topologies of these graphs are given in Appendix~\ref{ap1}.

\subsection{Experimental Setup \& Metrics}

The system setup was as follows: CPU: AMD Ryzen 9 3900X (12 cores); RAM: 32GB; Ubuntu 20.04. 

Each experiment is parameterized by the underlying graph \(\calG\), the number of agents \(n\), the number \(m\) of memory states per agent, the variance-punishing weight \(\kappa\), and the \(\alpha\) parameter weighting the value in case of agent failure. 
With an exception described later, the objective is to minimize the maximum over all vertices of the graph. 
I.e., for a given \(\kappa\) and \(\alpha\), the objective is
\begin{equation}\small
\label{obj-example-exps}
\begin{split}
  \textbf{minimize}\quad
		& \max\big\{ \MT(v,0) + \kappa\cdot\sqrt{\MTV(v,0)} \mid v \in V\big\}
			\\
  + & \alpha \cdot \max\big\{ \MT(v,1) + \kappa\cdot\sqrt{\MTV(v,1)} \mid v \in V\big\}.
\end{split}
\end{equation}
We use $\obj$ to denote the objective function of~\eqref{obj-example-exps}.

When \(\kappa\) or \(\alpha\) are nonzero, we report not only \(\obj\), but also the individual maxima of \(\MT^\mu(v,0)\), \(\MT^\mu(v,1)\), and \(\MTV^\mu(v,0)\) for the computed strategy \(\mu\), since these quantify how well \(\mu\) performs in case of agent (dis)functionality and what is its degree of stochastic stability:
\begin{align}
\label{eq:metrics}
\begin{split}
   \meanE &= \max\big\{ \MT^\mu(v,0) \mid v \in V \big\} \\
   \stdev &= \max\big\{ \sqrt{\MTV^\mu(v,0)} \mid v \in V \big\} \\
   \meanER & = \max\big\{ \MT^\mu(v,1) \mid v \in V \big\}
\end{split}
\end{align}

Apart from these metrics, we report the average time \(t\) per single optimization step of Algorithm~\ref{alg:optim}. All benchmarks were run with 600 steps. \(R\) is not reported if \(\alpha\) is 0.

In most of the experiments we compute \emph{coordinated strategies}. Experiments with independent strategies are marked with an asterisk. These experiments show that coordinated strategies perform better than uncoordinated ones and computing the latter does not yield any advantage in speed of synthesis.

For each experimental setup, we performed 5 runs with different seeds. We report the results of the run with the best \(\meanE\). The remaining details of the experimental configuration are given in Appendix \ref{ap1}.


\subsection{Experiments and Discussion}

In Experiment 1, we studied the \kpath{5} graph for which optimal values can be computed by hand (see Appendix), addressing questions (A) and (C). The setup and the results are presented in Table \ref{tab:5path}. We experimented with various levels of variance, resilience, and memory size.

The experiments demonstrate the phenomena discussed in Section~1. In particular, coordinated strategies with memory perform better than memoryless ones (line 1 vs. line 2) or uncoordinated ones (line 2 vs. line 7, where the strategy from Section~1 is found) and randomization helps, since the value is worse when randomness is penalized (line 2 vs. line 6). Also, there is a clear tradeoff between increased resilience and the ``optimistic'' \(\meanE\)-value which assumes that all agents work correctly (lines 2--5).
\begin{table}[t]
\small
\centering
\begin{tabular}{rrrcccc}
\toprule 
\multicolumn{3}{c}{Setup} & \multicolumn{4}{c}{Results} \\
\cmidrule(r){1-3} \cmidrule(l){4-7} 
\(m\) & $\kappa$ & $\alpha$ & $\meanE$ & $\stdev$ & $\meanER$ & t (s) \\
\cmidrule(r){1-3} \cmidrule(l){4-7} 
      1 &         0 &      0.00  & 2.72 &      1.30 &  N/A &  \(<\)0.01 \\
      3 &         0 &      0.00 & 2.00 &      1.00 &  N/A &  0.02 \\
      3 &         0 &      0.10 & 2.99 &      1.80 & 8.43 &  0.02 \\
      3 &         0 &      0.50 & 3.11 &      0.94 & 6.79 &  0.02 \\
      3 &         0 &      1.00 & 3.23 &      1.00 & 6.58 &  0.02 \\
      3 &         1 &      0.00 & 3.00 &      0.00 &  N/A &  0.02 \\
      \(*2\) &         0 &      0.00 & 2.44 &      1.16 &  N/A &  \(<\)0.01 \\
\bottomrule
\end{tabular}
\caption{Experiment 1: All benchmarks are \kpath{5} with 2 agents. Each row corresponds to a single instance of the experiment. The last line is an instance with \emph{uncoordinated agents.}}
\label{tab:5path}
\end{table}

\begin{table}[t]
\small
\centering
\begin{tabular}{rrccc}
\toprule 
\multicolumn{2}{c}{Setup} & \multicolumn{3}{c}{Results} \\
\cmidrule(r){1-2} \cmidrule(l){3-5} 
k & $\kappa$ & $\meanE$ & $\stdev$ & t (s) \\
\cmidrule(r){1-2} \cmidrule(l){3-5} 
7  &         0 &   4.01 &      1.97 &   0.05 \\
7  &         1 &   5.00 &      0.00 &   0.05 \\
9  &         0 &   5.85 &      3.06 &   0.17 \\
9  &         1 &   7.00 &      0.00 &   0.18 \\
11 &         0 &   7.76 &      4.47 &   0.61 \\
11 &         1 &   9.00 &      0.00 &   0.63 \\
13 &         0 &   9.92 &      5.02 &   1.71 \\
13 &         1 &  11.00 &      0.00 &   1.67 \\
$*7$  &         0 &  4.21 &      1.91 &   0.07 \\
$*9$  &         0 &  5.87 &      2.97 &   0.36 \\
\bottomrule
\end{tabular}

\caption{Experiment 2: Paths \kpath{k} for increasing odd values of \(k\) with 2 agents, 3 memory states per agent, and resilience parameter \(\alpha = 0\). The last two lines are with uncoordinated agents.}
\label{tab:paths}
\end{table}

In Experiment 2, we addressed question (B). We kept increasing the size of the open perimeter and observed the increase in runtime. We also experimented with the variance parameter to see whether the determinism/value payoff demonstrated on \kpath{5} can be observed also here. The results are presented in Table~\ref{tab:paths}. 

As expected, the runtime increases with graph size, although it stays well within practical limits. Penalizing randomness (even lines up to line 8) leads to deterministic strategies whose value is equivalent to the situation where the agents synchronously sweep the line as in Figure~\ref{fig-strat} (d). Allowing more randomness helps significantly. In case of \kpath{9}--\kpath{13}, the gain in \(\meanE\) is more than one, suggesting that the strategies perform more intricate behavior than simply splitting the line into 2 parts and sharing the middle node in a randomized way. Similar gain can be seen in the uncoordinated case, though it is not as large as in the coordinated one.


In Experiment 3, we analyzed two instances of the \(4\times 4\) grid graph with several removed edges. In G, the objective function is as described in~\eqref{obj-example-exps}. In H (which has a different topology than G), the maxima in \eqref{obj-example-exps} are only taken over a subset \(T \subseteq V\) of \emph{target} nodes. This simulates patrolling scenarios where \(T\) are the valuable targets to protect and the remaining nodes represent transit routes etc. We experiment with various resilience parameters to analyze the tradeoff between optimal values and resilience in this scenario, addressing questions (B)--(D).  The results are presented in Table~\ref{tab:grids}.

We can already observe a significant increase in computation time. The quality of the resulting strategy is affected by the graph topology and distribution of targets. In \(G\), already the optimization of \(\meanE\) leads to the best result and deterministic cycling through the graph. In \(H\), there is a tradeoff between \(\meanE\) and resilience, since in non-resilient case the agents are motivated to divide the target states among themselves. Again, randomization outperforms deterministic strategies (line 5 vs. line 6).
\begin{table}
\small
\centering
\begin{tabular}{crrcccc}
\toprule 
\multicolumn{3}{c}{Setup} & \multicolumn{4}{c}{Results} \\
\cmidrule(r){1-3} \cmidrule(l){4-7} 
$\calG$ & $\kappa$ & $\alpha$ & $\meanE$ & $\stdev$ & $\meanER$ & t (s) \\
\cmidrule(r){1-3} \cmidrule(l){4-7} 
$G$ &      0.00 &      0.10 & 13.60 &      8.38 & 27.52 & 19.58 \\
$G$ &      0.00 &      0.00 & 11.00 &      0.00 &   N/A &  6.12 \\
$G$ &      0.10 &      0.00 & 11.00 &      3.63 &   N/A &  6.11 \\
$H$ &      0.00 &      0.10 & 11.00 &      0.00 & 23.00 &  6.00 \\
$H$ &      0.00 &      0.00 & 10.86 &      6.12 &   N/A &  2.03 \\
$H$ &      0.10 &      0.00 & 11.00 &      0.00 &   N/A &  2.04 \\

\bottomrule
\end{tabular}
\caption{Experiment 3: All benchmarks are with 2 agents and three memory states per agent.}
\label{tab:grids}
\end{table}

\begin{table}
\small
\newcolumntype{M}{>{\centering\arraybackslash}p{0.7cm}}
\newcolumntype{N}{>{\centering\arraybackslash}p{1cm}}
\begin{tabular}{crrrMNNM}
\toprule 
\multicolumn{4}{c}{Setup} & \multicolumn{4}{c}{Results} \\
\cmidrule(r){1-4} \cmidrule(l){5-8} 
$\calG$ & \(m\) & $\kappa$ & $\alpha$ & $\meanE$ & $\stdev$ & $\meanER$ & t (s) \\
\cmidrule(r){1-4} \cmidrule(l){5-8} 
$\Delta$ &       2 &      0.00 &      0.00 & 2.16 &      1.11 &  N/A &  3.71 \\
$\Delta$ &       1 &      0.00 &      0.00 & 2.03 &      1.10 &  N/A &  0.69 \\
$P_5$    &       1 &      0.00 &      0.50 & 1.83 &      0.55 & 4.98 &  0.20 \\
$P_5$    &       1 &      0.10 &      0.10 & 1.00 &      0.00 & 5.04 &  0.18 \\
\bottomrule
\end{tabular}
\caption{Experiment 4: All benchmarks are with 3 agents.}
\label{tab:3a}
\end{table}

In Experiment 4, we focused on scenarios with three agents over the triangle and \kpath{5} graphs (questions (B)--(D)). In \trigr{}, the tool found a memoryless randomized strategy whose value is close to the optimal memoryless value.
On the \kpath{5} graph we see an interesting tradeoff between \( \meanE \) and resilience. Already for a relatively low \( \kappa \) the tool finds a 
deterministic strategy with finite \(R\) while keeping the optimistic value \(\meanE\) optimal. Further increase of \(\kappa\) leads to a modest improvement in resilience at the cost of a notable increase of \(\meanE\).


%% file: sections/conclusion.tex
\section{Conclusions}

Our results show that optimizing complex objectives involving expected time for visiting configurations, stochastic stability, and resilience is feasible for non-trivial instances. As it was mentioned in Section~\ref{sec-model}, the class of FTRV objectives can be further enriched by operators allowing for specifying \emph{long-run average} values of terms, and thus express other well-known performance measures such as average idleness. 

Another challenge is to improve the overall performance of our synthesis algorithm. Since the graphs representing environments are typically sparse, using appropriate data structure might lead to a substantial speedup. Another possibility is to reduce the complexity by identifying and exploiting various types of symmetries occuring in the analysis of the Markov chain $M_\mu$ for a given solution~$\mu$.

%% file: sections/graphs.tex
\begin{figure*}
	\begin{subfigure}[b]{0.36\textwidth}
		\centering
		\begin{tikzpicture}[x=2cm, y=2cm, scale=0.7,font=\small]
		\foreach \X in {0,1,2,3}{%
			 \foreach \Y in {0,1,2,3}{%
					\node [min] (\X\Y) at (\X,\Y) {};
			 }
		}
		\node [min,double] (00) at (0,0) {};
		\node [min,double] (03) at (0,3) {};
		\node [min,double] (13) at (1,3) {};
		\node [min,double] (33) at (3,3) {};
		\node [min,double] (32) at (3,2) {};
		\node [min,double] (30) at (3,0) {};
		\path [-] (02) edge (03);
		\path [-] (01) edge (00) edge (02) edge (11);
		\path [-] (11) edge (21) edge (10) edge (12);
		\path [-] (20) edge (10) edge (21);
		\path [-] (22) edge (21) edge (12) edge (23);
		\path [-] (23) edge (33) edge (13);
		\path [-] (31) edge (32) edge (30) edge (21);
		\end{tikzpicture}
		\caption{Graph $H$. Double circled nodes are the targets.}
	\end{subfigure}
	\hfill
	\begin{subfigure}[b]{0.36\textwidth}
		\centering
		\begin{tikzpicture}[x=2cm, y=2cm, scale=0.7,font=\small]
		\foreach \X in {0,1,2,3}{%
			 \foreach \Y in {0,1,2,3}{%
					\node [min] (\X\Y) at (\X,\Y) {};
			 }
		}
		\path [-] (12) edge (22) edge (13) edge (02);
		\path [-] (03) edge (13) edge (02);
		\path [-] (01) edge (02) edge (11) edge (00);
		\path [-] (10) edge (00) edge (11) edge (20);
		\path [-] (30) edge (20) edge (31);
		\path [-] (21) edge (20) edge (31);
		\path [-] (32) edge (31) edge (33);
		\path [-] (33) edge (23);
		\end{tikzpicture}
		\caption{Graph $G$.}
	\end{subfigure}
	\hfill
	\begin{subfigure}[b]{0.27\textwidth}
		\centering
		\begin{tikzpicture}[x=2cm, y=2cm, scale=0.7,font=\small]
		\node [min] (BL) at (0,0) {};
		\node [min] (BC) at (1.5,0) {};
		\node [min] (BR) at (3,0) {};
		\node [min] (T) at (1.5,2.55) {};
		\node [min] (TL) at (0.75,1.275) {};
		\node [min] (TR) at (2.25,1.275) {};
		\node [min] (C) at (1.5,0.85) {};
		\path [-] (C) edge (T) edge (BL) edge (BR);
		\path [-] (BC) edge (BL) edge (BR);
		\path [-] (TL) edge (BL) edge (T);
		\path [-] (TR) edge (BR) edge (T);
		\end{tikzpicture}
		\caption{Graph $\Delta$.}
	\end{subfigure}
	\caption{The structure of the graphs analyzed in our experiments.}	
	\label{fig:graphs}
\end{figure*}
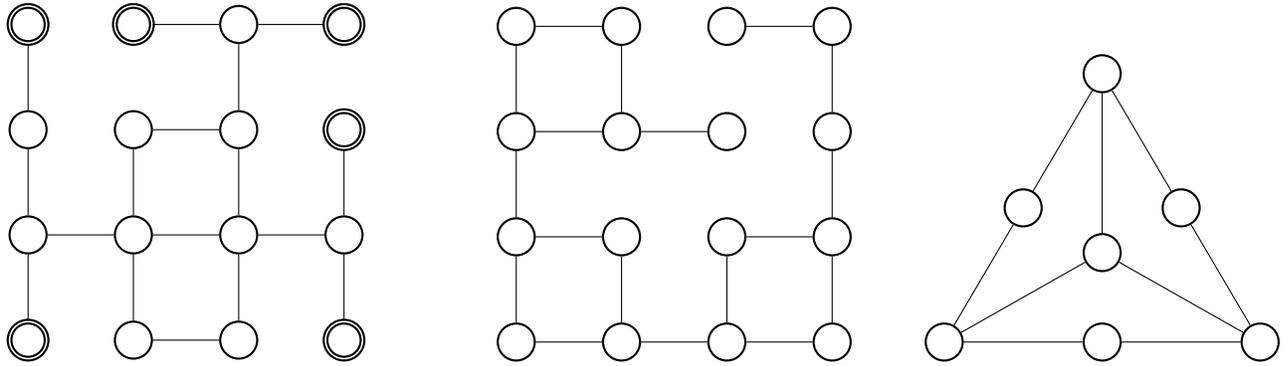

%% file: sections/appendix.tex
\section{Experiments and Source Code}
\label{ap1}

The structure of the graphs $G$, $H$, and $\Delta$ from our experiments are shown in Fig.~\ref{fig:graphs}.
The code and experiments setup can be found at

\smallskip
\centerline{\url{https://gitlab.fi.muni.cz/formela/2023-ijcai-multi-agents}.}

\section{Proofs}

We prove that coordinated agents are strictly stronger than autonomous agents. In particular, recall the example from Section 1 (path on vertices $V=\{A,B,C,D,E\}$ with 2 agents).
We have shown that there is a strategy for coordinated agents whose value of the objective function $E \equiv \max\{\MT(v,0) \mid v \in V\}$ is $2$. Now, we show that for every strategy for autonomous agents, the value of $E$ is greater than $2$.

For the sake of contradiction, assume that there is a strategy for autonomous agents whose value of $E$ is at most $2$. This means that for every reachable configuration $c$ and every vertex $v\in V$, the expected time of visiting $v$ from $c$ is at most $2$. For $K,L\in V$, we use $KL$ to denote a configuration where one agent is in $K$ and the other agent is in $L$. The idea of the proof is as follows: We show that the configurations $CC$, $BC$ and $AC$ must be unreachable. By the symmetry between $B$ and $D$, and also between $A$ and $E$, this implies that the configurations $DC$ and $EC$ are also unreachable. Hence, $C$ is unreachable, which is a contradiction.

The easiest case is $CC$. Thus, assume that configuration $CC$ is reached at (say) time $0$. Since $A$ must be visited from $CC$ in expectedly at most $2$ steps and it is not visited at time $0$ and cannot be visited at time $1$, it must be visited at time $2$ with probability $1$. The same argument can be used for $E$. Therefore, the configurations at times $1,2,3$ must be $BD,AE,BD$, respectively. Then, from $BD$ at time $1$, $C$ is not visited within $2$ steps at all.

Now, we rule out $BC$. Again, assume that configuration $BC$ is reached at time $0$.
Let $X$ be the agent in $B$ and $Y$ be the agent in $C$. Similarly as above, it can be shown that $Y$ must go via $D$ to $E$ in the next $2$ steps, otherwise $E$ would not be visited in expectedly at most $2$ steps. In particular, we have that $X$ is in $B$ at time $0$ and $Y$ is neither in $A$ nor in $C$ at times $1,2,3$. From $B$ at time $0$, $X$ must go to $A$ with positive probability, otherwise $A$ would not be visited within $2$ steps at all. Therefore, at time $1$, it is possible that the configuration is $AD$ with $Y$ going to $E$ with probability $1$ in the next step. From this configuration, $X$ must go via $B$ to $C$ in the next $2$ steps, otherwise $C$ would not be visited in expectedly at most $2$ steps (again, $C$ is visited neither at time $1$ nor at time $2$, so it must be visited at time $3$ with probability $1$). Then, from $BE$ at time $2$, $A$ is not visited within $2$ steps at all.

Finally, assume that configuration $AC$ is reached at time $0$. Again, let $X$ be the agent in $A$ and $Y$ be the agent in $C$. As above, we know that $Y$ must go via $D$ to $E$ in the next $2$ steps. Then, $Y$ goes to $D$ and, at time $4$, $Y$ may be located in either $C$ or $E$. Assume that $Y$ is in $E$ at time $4$ with probability $1$. Then, we have that $X$ is in $B$ at time $1$ and $Y$ is neither in $A$ nor in $C$ at times $2,3,4$. This is the same situation (shifted by $1$ time unit) as the one from the $BC$ case above, which we have already refuted. Therefore, $Y$ must be in $C$ at time $4$ with positive probability. Since we have already shown that configuration $CC$ must be unreachable, we get that $X$ must not be in $C$ at time $4$.\footnote{Note that this is the (only) place in the proof where we use the fact that the agents are autonomous. For coordinated agents, $X$ could be in $C$ provided $Y$ would be in $E$. However, an autonomous agent's moves are independent of the other agent's moves. Therefore, since $Y$ might be in $C$ at time $4$ and $CC$ must be unreachable, it follows that the probability of $X$ being in $C$ at time $4$ must be $0$.} Furthermore, we know that from $B$ at time $1$, $X$ must go to $A$ with positive probability, otherwise $A$ would not be visited within $2$ steps at all.
Therefore, at time $2$, it is possible that the configuration is $AE$, followed by $BD$ at time $3$. Since we have already shown that $X$ cannot be in $C$ at time $4$, it must be the case that $Y$ is in $C$ at time $4$ with probability $1$, otherwise from $AE$ at time $2$, $C$ would not be visited in expectedly at most $2$ steps.
Then, from $BD$ at time $3$, $E$ is not visited within $2$ steps at all, and we are done.

%% file: arxiv.bbl
\begin{thebibliography}{}

\bibitem[\protect\citeauthoryear{Almeida \bgroup \em et al.\egroup
  }{2004}]{ARSTMCC:multi-patrolling-survey}
A.~Almeida, G.~Ramalho, H.~Santana, P.~Tedesco, T.~Menezes, V.~Corruble, and
  Y.~Chevaleyr.
\newblock Recent advances on multi-agent patrolling.
\newblock {\em Advances in Artificial Intelligence -- SBIA}, 3171:474--483,
  2004.

\bibitem[\protect\citeauthoryear{Basilico \bgroup \em et al.\egroup
  }{2016}]{BLM:Remote-SW-Protection}
N.~Basilico, A.~Lanzi, and M.~Monga.
\newblock A security game model for remote software protection.
\newblock In {\em Proceedings of {ARES} 2016}, pages 437--443, 2016.

\bibitem[\protect\citeauthoryear{Br{\'{a}}zdil \bgroup \em et al.\egroup
  }{2017}]{BCFK:performance-stability-JCSS}
T.~Br{\'{a}}zdil, K.~Chatterjee, V.~Forejt, and A.~Ku{\v{c}}era.
\newblock Trading performance for stability in {Markov} decision processes.
\newblock {\em Journal of Computer and System Sciences}, 84:144--170, 2017.

\bibitem[\protect\citeauthoryear{Ceccato and
  Tonella}{2011}]{CT:Codebender-IEEESoftware}
M.~Ceccato and P.~Tonella.
\newblock Codebender: Remote software protection using orthogonal replacement.
\newblock {\em IEEE Software}, 28(2):28--34, 2011.

\bibitem[\protect\citeauthoryear{Choset}{2005}]{Choset:book}
H.M.{} Choset.
\newblock {\em Principles of Robot Motion: Theory, Algorithms, and
  Implementation}.
\newblock MIT Press, 2005.

\bibitem[\protect\citeauthoryear{Collberg \bgroup \em et al.\egroup
  }{2012}]{CMMN:distributed-tamper-detection}
C.~Collberg, S.~Martin, J.~Myers, and J.~Nagra.
\newblock Distributed application tamper detection via continuous software
  updates.
\newblock In {\em Proceedings of the 28th Annual Computer Security Applications
  Conference}, pages 319--328. ACM Press, 2012.

\bibitem[\protect\citeauthoryear{Czyzowicz \bgroup \em et al.\egroup
  }{2017}]{CGKKKT:corrupted-patrolmen-Algorithmica}
J.~Czyzowicz, L.~Gasieniec, A.~Kosowski, E.~Kranakis, D.~Krizanc, and N.~Taleb.
\newblock When patrolmen become corrupted: Monitoring a graph using faulty
  mobile robots.
\newblock {\em Algorithmica}, 79:925--940, 2017.

\bibitem[\protect\citeauthoryear{Dixon}{2019}]{Dixon:book}
A.~Dixon.
\newblock {\em Multi-Agent Systems: Design, Synthesis and Analysis}.
\newblock Clanrye International, 2019.

\bibitem[\protect\citeauthoryear{Gronauer and
  Diepold}{2022}]{GD:multiagent-learning-survey}
S.~Gronauer and K.~Diepold.
\newblock Multi-agent deep reinforcement learning: A survey.
\newblock {\em Artificial Intelligence Review}, 55:895--943, 2022.

\bibitem[\protect\citeauthoryear{Hazon and
  Kaminka}{2005}]{HK:robustness-multirobot-ICRA}
N.~Hazon and G.A. Kaminka.
\newblock Redundancy, efficiency and robustness in multi-robot coverage.
\newblock In {\em Proceedings of ICRA 2005}, pages 735--741. IEEE Computer
  Society Press, 2005.

\bibitem[\protect\citeauthoryear{Huang \bgroup \em et al.\egroup
  }{2019}]{HZHH:multi-robot-patrol-survey}
L.~Huang, M.~Zhou, K.~Hao, and E.~Hou.
\newblock A survey of multi-robot regular and adversarial patrolling.
\newblock {\em IEEE/CAA Journal of Automatica Sinica}, 6(4):894--903, 2019.

\bibitem[\protect\citeauthoryear{Kawamura and
  Soejima}{2020}]{KS:Multi-patrol-strategies-TCS}
A.~Kawamura and M.~Soejima.
\newblock Simple strategies versus optimal schedules in multi-agent patrolling.
\newblock {\em Theoretical Computer Science}, 839:195--206, 2020.

\bibitem[\protect\citeauthoryear{Kingma and Ba}{2015}]{Adam}
D.~P. Kingma and J.~Ba.
\newblock Adam: A method for stochastic optimization.
\newblock In {\em Proceedings of {ICLR} 2015}, 2015.

\bibitem[\protect\citeauthoryear{Kla{\v{s}}ka \bgroup \em et al.\egroup
  }{2021}]{KKMR:Regstar-UAI}
D.~Kla{\v{s}}ka, A.~Ku{\v{c}}era, V.~Musil, and V.~{\v{R}}eh{\'{a}}k.
\newblock Regstar: Efficient strategy synthesis for adversarial patrolling
  games.
\newblock In {\em Proceedings of UAI 2021}, pages 471--481, 2021.

\bibitem[\protect\citeauthoryear{Kla{\v{s}}ka \bgroup \em et al.\egroup
  }{2022}]{KKMR:RecReach-IJCAI}
D.~Kla{\v{s}}ka, A.~Ku{\v{c}}era, V.~Musil, and V.~{\v{R}}eh{\'{a}}k.
\newblock General optimization framework for recurrent reachability objectives.
\newblock In {\em Proceedings of the International Joint Conference on
  Artificial Intelligence (IJCAI-ECAI 2022)}, pages 4642--4648, 2022.

\bibitem[\protect\citeauthoryear{LaValle}{2006}]{LaValle:book}
S.M.{} LaValle.
\newblock {\em Planning Algorithms}.
\newblock Cambridge University Press, 2006.

\bibitem[\protect\citeauthoryear{Norris}{1998}]{Norris:book}
J.R.{} Norris.
\newblock {\em {Markov} Chains}.
\newblock Cambridge University Press, 1998.

\bibitem[\protect\citeauthoryear{Paszke \bgroup \em et al.\egroup
	}{2019}]{PyTorch}
A.{} Paszke, S.{} Gross, F.{} Massa, A.{} Lerer, J.{} Bradbury, G.{}
	Chanan, T.{} Killeen, Z.{} Lin, N.{} Gimelshein, L.{} Antiga, A.{}
	Desmaison, A.{} Kopf, E.{} Yang, Z.{} DeVito, M.{} Raison, A.{}
	Tejani, S.{} Chilamkurthy, B.{} Steiner, Lu~Fang, J.{} Bai, and S.{}
	Chintala.
\newblock Pytorch: An imperative style, high-performance deep learning library.
\newblock In {\em Advances in Neural Information Processing Systems 32}, pages
	8024--8035. Curran Associates, Inc., 2019.

\bibitem[\protect\citeauthoryear{Portugal and
  Rocha}{2011}]{PR:multi-patrolling-survey}
D.~Portugal and R.~Rocha.
\newblock A survey on multi-robot patrolling algorithms.
\newblock {\em Technological Innovation for Sustainability}, 349:139--146,
  2011.

\bibitem[\protect\citeauthoryear{Shoham}{2008}]{Shoham:book}
Y.~Shoham.
\newblock {\em Multiagent Systems}.
\newblock Cambridge University Press, 2008.

\bibitem[\protect\citeauthoryear{Sinha \bgroup \em et al.\egroup
  }{2018}]{SFAKT:Stackelberg-Security-Games}
A.~Sinha, F.~Fang, B.~An, C.~Kiekintveld, and M.~Tambe.
\newblock Stackelberg security games: Looking beyond a decade of success.
\newblock In {\em Proceedings of the International Joint Conference on
  Artificial Intelligence (IJCAI 2018)}, pages 5494--5501, 2018.

\bibitem[\protect\citeauthoryear{Smith \bgroup \em et al.\egroup
  }{2011}]{STBR:temporal-planning-IJRS}
S.L.{} Smith, J.~T{\r{u}}mov{\'{a}}, C.~Belta, and D.~Rus.
\newblock Optimal path planning for surveillance with temporal-logic
  constraints.
\newblock {\em International Journal of Robotics Research}, 30(14):1695--1708,
  2011.

\bibitem[\protect\citeauthoryear{Tarjan}{1972}]{Tarjan:SCC-decomp-SICOMP}
R.~Tarjan.
\newblock Depth-first search and linear graph algorithms.
\newblock {\em {SIAM} Journal of Computing}, 1(2), 1972.

\bibitem[\protect\citeauthoryear{Toth and Vigo}{2001}]{TV:book}
P.~Toth and D.~Vigo.
\newblock {\em The Vehicle Routing Problem}.
\newblock SIAM Monographs on Discrete Mathematics and Applications. SIAM, 2001.

\bibitem[\protect\citeauthoryear{van~der Hoek and
  Wooldridge}{2012}]{HW:logics-multiagent-survey}
W.~van~der Hoek and M.~Wooldridge.
\newblock Logics for multiagent systems.
\newblock {\em AI Magazine}, 33(3):92--105, 2012.

\bibitem[\protect\citeauthoryear{Wooldridge}{2009}]{Wooldridge:book}
M.~Wooldridge.
\newblock {\em Introduction to {MultiAgent} Systems}.
\newblock Wiley, 2009.

\bibitem[\protect\citeauthoryear{Yin \bgroup \em et al.\egroup
  }{2010}]{YKKCT:Stackelberg-Nash-security}
Z.~Yin, D.~Korzhyk, C.~Kiekintveld, V.~Conitzer, and M.~Tambe.
\newblock Stackelberg vs.{} {Nash} in security games: Interchangeability,
  equivalence, and uniqueness.
\newblock In {\em Proceedings of AAMAS 2010}, pages 1139--1146, 2010.

\end{thebibliography}
